\documentstyle[prl,aps,epsf]{revtex}
\begin{document}
\twocolumn[
\hsize\textwidth\columnwidth\hsize\csname@twocolumnfalse\endcsname
%*******************************************************************************
%     Header
%*******************************************************************************
\title
{
Dephasing by time-dependent random potentials
}

\author{Takeshi Nakanishi}
\address
{
Electrotechnical Laboratory, 1-1-4 Umezono, Tsukuba, Ibaraki 305, Japan
}
\author{Tomi Ohtsuki}
\address
{
Department of Physics, Sophia University, Kioi-cho 7-1, Chiyoda-ku,
Tokyo 102, Japan
}
\author{Tohru Kawarabayashi}
\address
{
Institute for Solid State Physics, University of Tokyo,
Roppongi, Minato-ku, Tokyo 106, Japan
}
\date{\today}
\maketitle
%*******************************************************************************
%     Abstruct
%*******************************************************************************
\begin{abstract}
Diffusion of electrons in a two-dimensional system with time-dependent
random potentials is investigated numerically.
The correction to the conductivity due to inelastic scatterings
by oscillating potentials is shown to be a universal function of 
the frequency $\omega$, which is consistent with  the weak
localization prediction $\frac{e^2}{3\pi^2 \hbar}\log \omega$.
\end{abstract}
]
%\kword{weak localization, equation of motion method}
%*******************************************************************************
%     Introduction
%*******************************************************************************
Transport properties of the two-dimensional disordered electron systems have
attracted much attention in recent years.
To understand the transport properties of a random system, the
concept of quantum interference plays an important
role.
The constructive interference of two time-reversed
trajectories enhances the probability for backscattering,
leading to the contribution called
weak localization correction.\cite{abrahams79,gorkov79}
Another example is the phenomenon of the sample specific reproducible
conductance fluctuation,\cite{lee85,altshuler85}
which is highly sensitive
to the motion of a single impurity atom in
the metallic region.\cite{feng86,feng93,klepper91,beulter87}
More recently, such impurity-motion-induced conductance fluctuations
have been utilized experimentally to study the detailed dynamics of
impurity atom tunneling in a single two-level system in disordered
metals.\cite{zimmerman91}
The motion of an interstitial impurity between two energetically
equivalent stability points that have different effects on the conductivity
is also studied to understand the temporal fluctuations of the conductance,
where one can observe $1/f$ noise.\cite{weissman88}

Such effects of quantum interference are suppressed in the
presence of the inelastic scattering, for instance, due to
the electron-phonon interaction or due to the Coulomb interaction
of conduction electrons. There exists a time scale called dephasing
time $\tau_{\phi}$, after which the interference effects are
destroyed.\cite{altshuler80,fukuyama80}
The dephasing time is determined by the inelastic process and usually
depends on the magnetic field or on the temperature.
Consequently, the effect of interference can be observed experimentally
by controlling these parameters .

In this paper, we study the effect of dynamically fluctuating potentials
on the conductivity by solving the time-dependent Schr\"odinger
equation numerically.
We adopt here the formula for solving the time-dependent Schr\"odinger
equation proposed
by Suzuki,\cite{suzuki90,suzuki91,suzuki92,suzuki92b,suzuki93,suzuki94}
which make it possible to deal numerically with the time-dependent potential
on the sufficiently large system of two-dimensional lattice.
We observe the second moment of the wave
packet, which gives information about extension of the wave packet.
We then evaluate the conductivity and discuss how the
conductivity depends on the motion of potentials, in particular, on
the frequency of potentials.

%*******************************************************************************
%     Model & Method
%*******************************************************************************
In order to solve numerically the time-dependent Schr\"odinger
equations, we adopt the method based on the higher-order
decomposition of exponential operators.\cite{suzuki90,raedt87}
The basic formula we have used is the fourth-order decomposition of
exponential operators:
\begin{eqnarray}
& &  \exp[x(A_1\ +\ A_2\ +\ \cdots\ +\ A_q)]\nonumber\\
& = & S(xp)S\left(x(1-2p)\right)S(xp)
 +  O(x^5),
\label{eqn:4decomposition}
\end{eqnarray}
where
\begin{equation}
S(x)\ =\ e^{xA_1/2}e^{xA_2/2}\cdots e^{xA_{q-1}/2}e^{xA_q}e^{xA_{q-1}/2}
\cdots e^{xA_1/2},
\end{equation}
and the parameter $p$ is given by $p\ =\ (2-\sqrt[3]{2})^{-1}$.
Here $A_1,\dots,A_q$ are arbitrary operators.

We consider the tight-binding Hamiltonian with time-dependent
potential $\varepsilon_i (t)$ on the two-dimensional square lattice:
\begin{equation}
H(t)\ =\ -\ V\sum_{(i,j)} C_i^{\dagger} C_j\
+\ \sum_i \varepsilon_i (t)C_i^{\dagger} C_i,
\label{eqn:hamiltonian}
\end{equation}
where $C^{\dagger}_{i}(C_{i})$ denotes a creation (annihilation)
operator of an electron at the site $i$ and $(i,j)$ is the nearest
neighbor site.
We then decompose this Hamiltonian into five parts as described in ref.
\cite{kawarabayashi95}, namely,
\begin{eqnarray}
H(t) & = & \sum_{n=1}^{4} H_n\ +\ H_{5}(t),\\
H_1 & \equiv &  V \sum_{r_x \in {\rm odd}} \sum_{r_y}
C_{\vec{r}+\hat{x}}^{\dagger}
C_{\vec{r}} + {\rm h. c.},\nonumber\\
H_2 & \equiv & V \sum_{r_x \in {\rm even}} \sum_{r_y}
C_{\vec{r}+\hat{x}}^{\dagger}
C_{\vec{r}} + {\rm h. c.},\nonumber\\
H_3 & \equiv & V \sum_{r_x} \sum_{r_y \in {\rm odd}}
C_{\vec{r}+\hat{y}}^{\dagger}
C_{\vec{r}} + {\rm h. c.},\nonumber\\
H_4 & \equiv & V \sum_{r_x} \sum_{r_y \in {\rm even}}
C_{\vec{r}+\hat{y}}^{\dagger}
C_{\vec{r}} + {\rm h. c.},\nonumber\\
H_{5}(t) & \equiv & \sum_{\vec{r}} \varepsilon_{\vec{r}}(t)
C_{\vec{r}}^{\dagger}
C_{\vec{r}} ,\label{eqn:hamiltonian5}
\end{eqnarray}
where $\hat{x}(\hat{y})$ denotes the unit vector in the $x(y)-$direction
and $r_x(r_y)$ the $x(y)$ component
of the position vector $\vec{r}$ of the sites.
All the length-scales are measured in units of the lattice constant $a_0$.
Each Hamiltonian $H_n$ consists of operators which commute with each other,
hence we can obtain the analytical expressions
for $\exp(-{\rm i}H_n t/\hbar)$ by diagonalizing two by two matrices.

The state vector $|\psi(t\ +\ \delta t)\rangle$ at time $t\ +\ \delta t$
is obtained as
\begin{equation}
|\psi(t\ +\ \delta t)\rangle\ =\ U(t\ +\ \delta t,t)|\psi(t)\rangle,
\label{eqn:timedevelopment}
\end{equation}
where the time-development operator $U(t\ +\ \delta t,t)$ is defined by
\begin{equation}
U(t\ +\ \delta t,t)\ =\ T\left(\exp{\left[-\frac{i}{\hbar}
\int_{t}^{t\ +\ \delta t}H(s)ds\right]}\right),
\label{eqn:developmentop}
\end{equation}
with $T$ the time-ordering operator.
Using the formula proposed by Suzuki,\cite{suzuki93} this ordered
exponential can be expressed by an ordinary exponential operator as
\begin{equation}
U(t\ +\ \delta t,t)\ =\ \exp{\left[\delta t(-\frac{i}{\hbar}H(t)\
+\ {\cal T})\right]},
\label{eqn:exptime}
\end{equation}
where the super-operator $\cal T$ is defined by
\begin{equation}
F(t)e^{\delta t{\cal T}}G(t)\ =\ F(t\ +\ \delta t)G(t).
\end{equation}
Using formula (\ref{eqn:4decomposition}) and
(\ref{eqn:exptime}),
the time-development operator (\ref{eqn:developmentop}) is decomposed
as the product of exponential operators;\cite{suzuki93}
\begin{eqnarray}
U(t+\delta t,t)\ = & & S_2(-{\rm i}\delta t p/\hbar,\ t +(1-p/2)\delta
t)\nonumber\\
& \times & S_2(-{\rm i}\delta t(1-2p)/\hbar,\ t +\delta t/2)\nonumber\\
& \times & S_2(-{\rm i}\delta t p/\hbar,\ t +p\delta t/2)\nonumber\\
& + &  O(\delta t^5)
\end{eqnarray}
with
\begin{eqnarray}
 S_2(x,t) & \equiv & {\rm e}^{xH_1/2}{\rm e}^{xH_2/2}\cdots
{\rm e}^{xH_4/2} {\rm e}^{xH_5(t)}\nonumber\\
& \times& {\rm e}^{xH_4/2} \cdots
{\rm e}^{xH_2/2}{\rm e}^{xH_1/2}.
\end{eqnarray}

In order to consider the time evolution of the wave packet with  fixed
energy $E$, we have carried out a numerical diagonalization  of
a subsystem $H_{N_0}$ whose size is $N_0$ by $N_0$, located
at the center of the whole system and have chosen an eigenstate
of $H_{N_0}$ with eigenvalue $E_{N_0} \approx E$ as the initial wave packet.

The quantity we observe is the second moment of the wave
packet $\langle r^2 (t) \rangle_c $ defined by
\begin{equation}
\langle r^2 (t) \rangle_c \equiv \langle |\vec{r} (t)|^2
\rangle -
\langle \vec{r}(t)\rangle^2
\end{equation}
with
\begin{equation}
\langle r^2 (t) \rangle \equiv \int {\rm d}\Omega r^{d-1} {\rm d}r
r^2 |\psi(\vec{r},t)|^2,
\end{equation}
where $\psi(\vec{r},t)$ denotes the wave function at time $t$,
and $d$ the dimensionality of the system.
If the wave function extends throughout the whole system,
the second moment is expected to grow in proportion to time $t$
\begin{equation}
\langle r^2 \rangle_c\ =\ 2dDt.
\label{eqn:diffusion}
\end{equation}
Here, the coefficient $D$ denotes the diffusion coefficient.
In contrast, if the wave function is localized, it is clear that the second
moment remains finite in the limit $t \rightarrow \infty$.\cite{raedt87}
In the metallic region the Einstein relation
\begin{equation}
\sigma\ =\ e^2 \rho (E_{\mbox{F}})D
\label{eqn:einstein}
\end{equation}
relates the conductivity $\sigma$ to the diffusion
constant, where $\rho (E_{\mbox{F}})$ is the density of state at Fermi
energy $E_{\mbox{F}}$.

%*******************************************************************************
%     Results
%*******************************************************************************
In order to take into account the effect of moving potential,
we assume that the site-potentials take the form:
\begin{equation}
\varepsilon_{\vec{r}}(t)\ =\ \varepsilon_{\vec{r}}(0) \cos{\left[\omega t +
\theta\left(\vec{r}\right)\right]},
\label{eqn:modelpotential}
\end{equation}
where $\omega$ is the frequency.
Effects of scattering from impurities are introduced through
randomness of  site energy $\varepsilon_{\vec{r}}(0)$ and phase
$\theta(\vec{r})$ at $t\ =\ 0$ distributed uniformly in the regions,
\begin{eqnarray}
|\varepsilon_{\vec{r}}(0)| & < & W/2,\nonumber\\
|\theta(\vec{r})| & < & \pi.
\label{eqn:randomnessrange}
\end{eqnarray}
We consider the adiabatic case $\omega \ll V/\hbar$, where impurities
move slower than electrons.

We have calculated the second moment
of the wave packet $\langle r^2 \rangle_c$
at various random potential strength.
The size of the systems are 500 by 500 for the energies
$E_{\mbox{F}}/V\ =\ -1$.
We have carried out an exact diagonalization for the 20 by 20 subsystem
at the center of the system and taken the eigenfunction of the subsystem
whose eigenvalue is closest to the given energy $E_{\mbox{F}}$
as the initial wave packet.
By this procedure we can simulate the diffusion of the wave
packet whose energy is approximately equal to $E_{\mbox{F}}$.
The single time step $\delta t$ is taken to be $0.2 (\hbar/V)$
in the simulation.
With this choice of $\delta t$, fluctuations of the
expectation value of the Hamiltonian
$\langle H \rangle \equiv \langle
\psi (t) | H | \psi (t) \rangle
$
for $\omega=0$ can be safely neglected~\cite{raedt87,kawarabayashi95}
throughout our simulations ($ t \leq 2000 (\hbar/V)$).

\begin{figure}[t]
\epsfxsize8cm
\epsffile{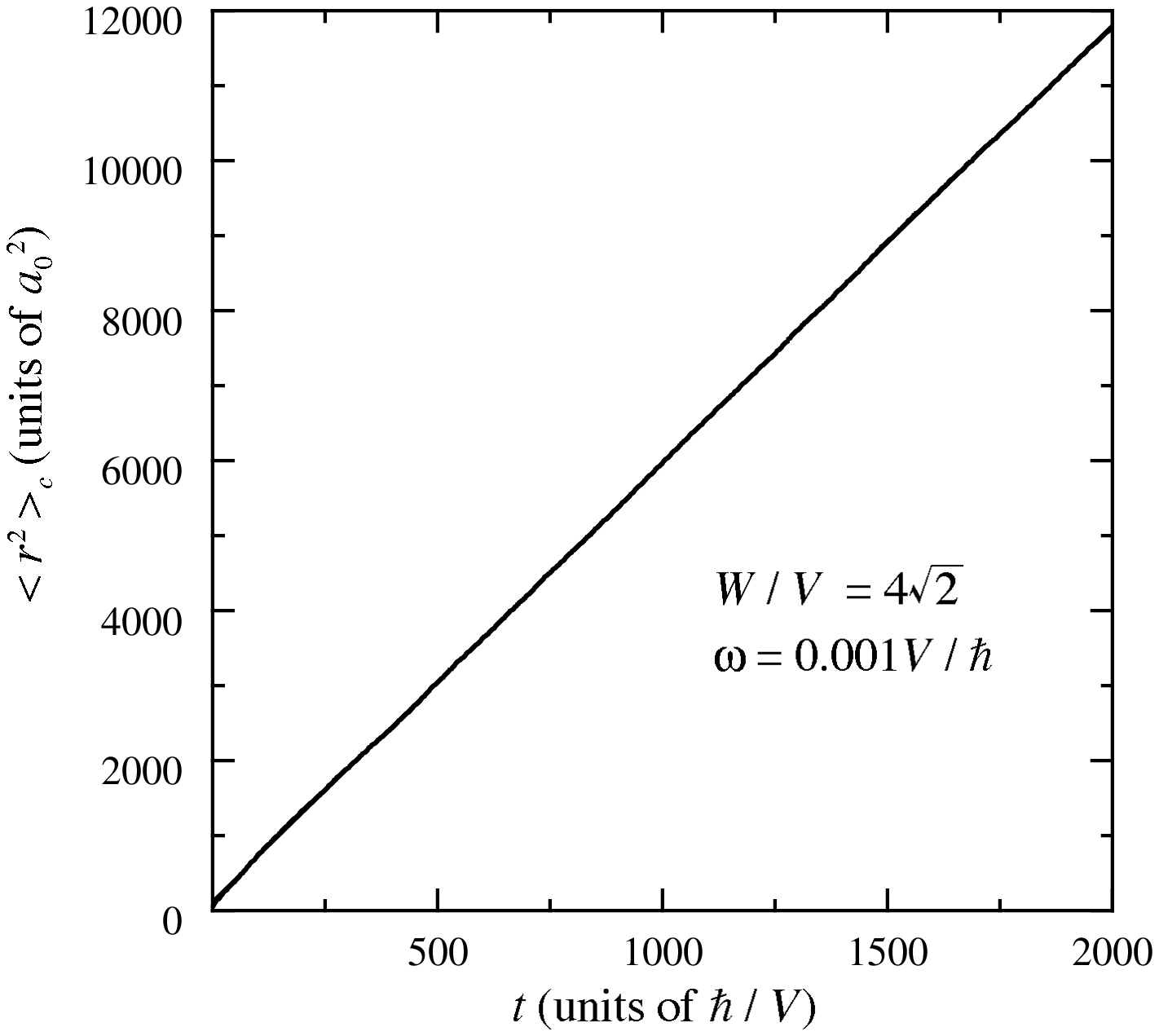}
\caption{\label{fig:1sample}}
\end{figure}
Fig. \ref{fig:1sample} gives an example of the calculated
second moment for $\omega\ =\ 10^{-3} V/\hbar,$ $W/V\ =\ 4\sqrt{2}$.
The second moment is proportional to the time
in a wide range up to $2\times10^{3} \hbar/V$.
We can evaluate the diffusion constant of this system from Eq.
(\ref{eqn:diffusion}) with least square fit to these data
and obtain the diffusion constant of the sample to be $D\ =\ 1.46a^2 V/\hbar$.
In the actual simulation, the quantities $\langle r^2 \rangle_c$ are
averaged over at least  five samples of random potential distribution.
The density of states $\rho (E_{\mbox{F}})$ in Eq.
(\ref{eqn:einstein}) is evaluated from the direct diagonalization
of the two-dimensional lattice model for 40 by 40 sites for
 $\omega\ =\ 0$,
and an average over 100 samples is performed.
We have obtained that
$\rho (E_{\mbox{F}})a_{0}^{2}V\ =\ 0.292,\ 0.291,\
0.281,$ and $0.251$ for $W/V\ =\ 2,\ 2\sqrt{2},\ 4,$ and $4\sqrt{2}$,
respectively.

\begin{figure}[b]
\epsfxsize8cm
\epsffile{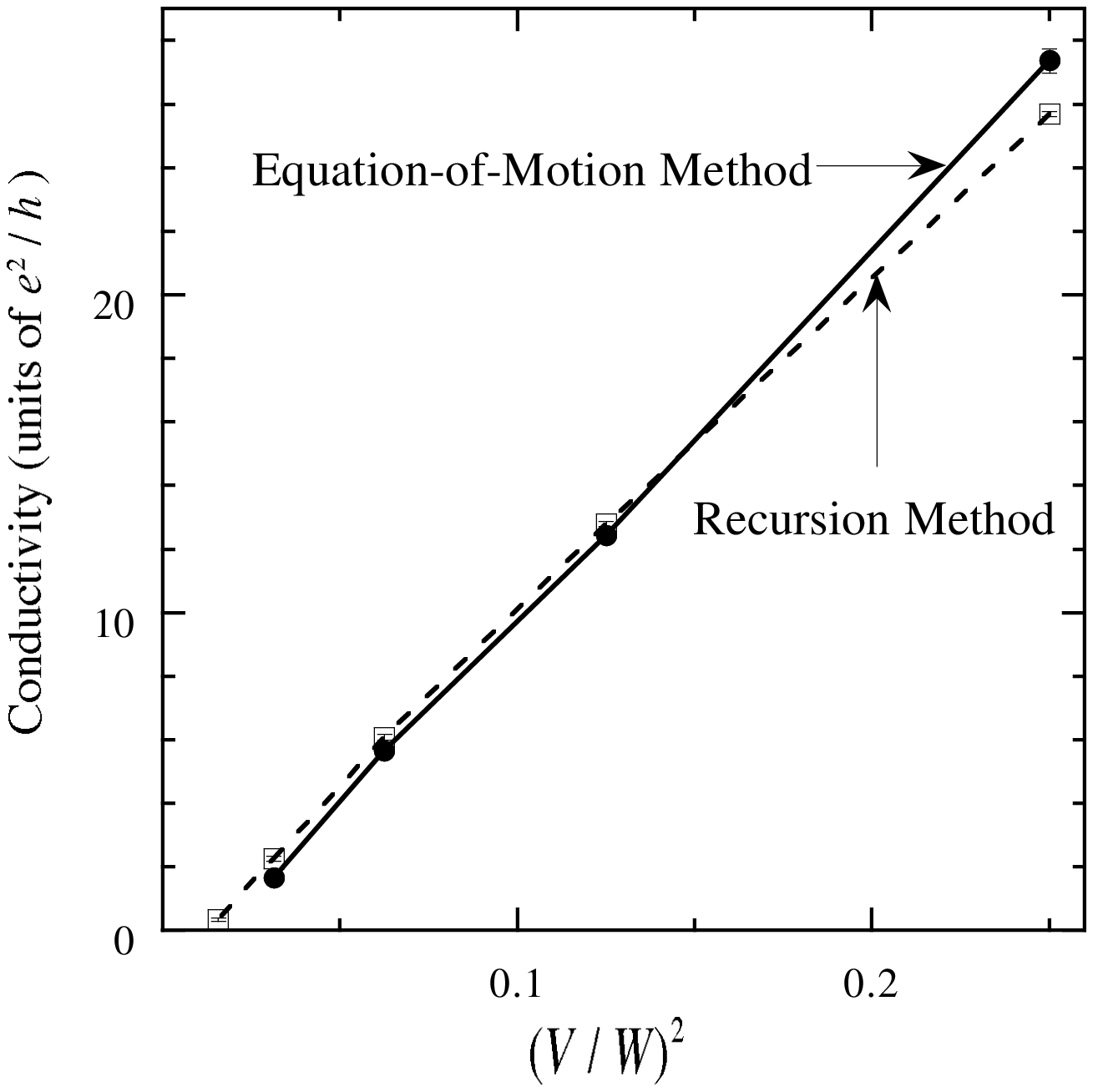}
\caption{\label{fig:recursion}}
\end{figure}
In Fig. \ref{fig:recursion}, we show the conductance for
$\omega\ =\ 0$ obtained by the present method, and compare it
with that
calculated by a conventional recursive Green's function method.\cite{ando91}
The results show that two methods agree fairly well with each other.
We can also see the conductivity is proportional to $(W/V)^{-2}$ in
a wide range of $W/V$.

\begin{figure}[t]
\epsfxsize8cm
\epsffile{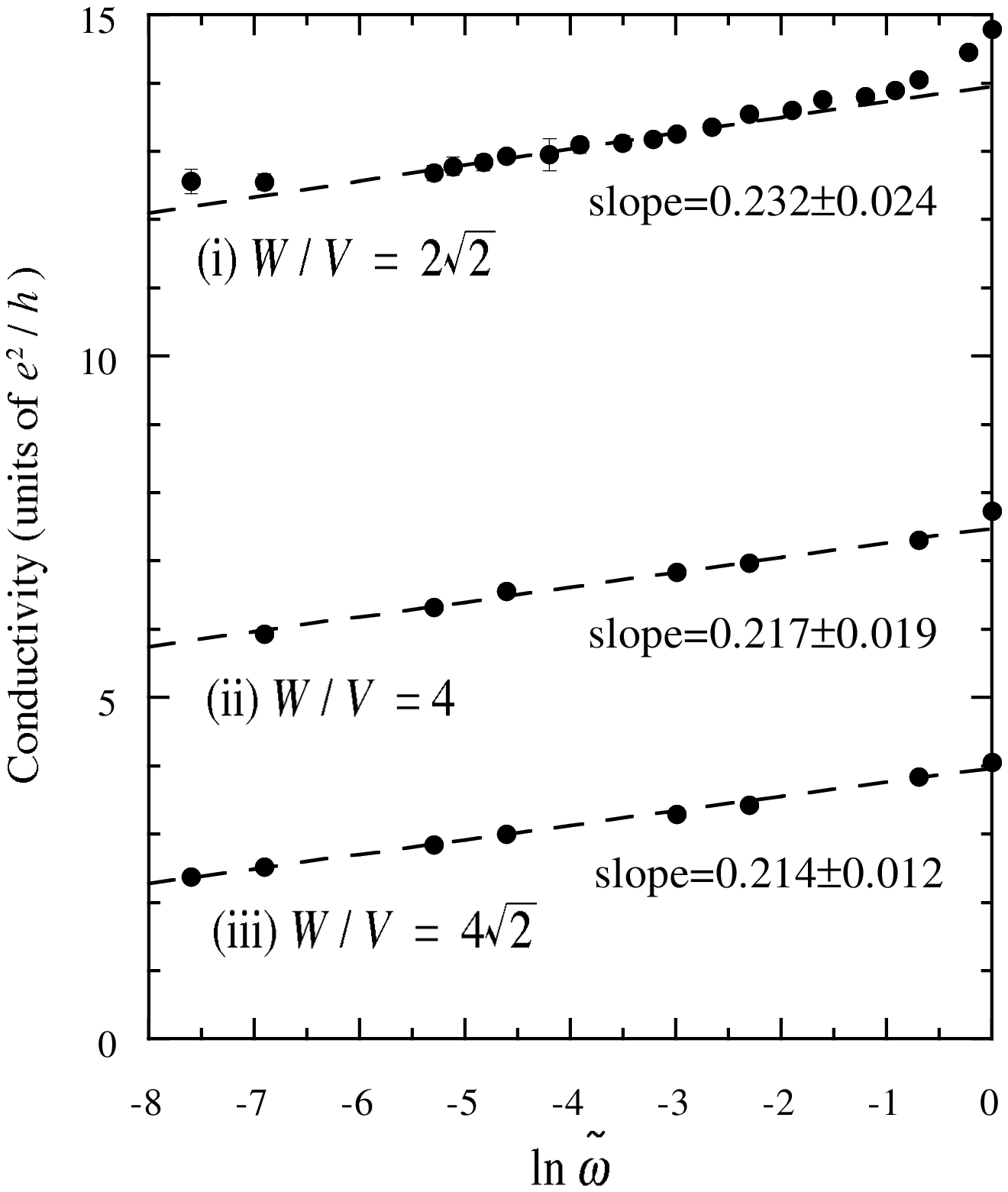}
\caption{\label{fig:universal}}
\end{figure}
In Fig. \ref{fig:universal} we show examples of the conductivity
as the function of $\ln{\tilde{\omega}}$ for $W/V\ =\ 2\sqrt{2}, 4,$
and $4\sqrt{2}$,
where $\tilde{\omega}\ =\ \omega/(V/\hbar)$.
The conductivity linearly depends on $\ln{\tilde{\omega}}$ as
\begin{equation}
\sigma\ =\ A\ln{\tilde{\omega}}\ +\ \sigma(0) .
\end{equation}
The coefficients $A(\approx 0.22)$
calculated by the least square fit
to the averaged conductivity, are almost the same among these three
different disorder cases, although the conductivity
$\sigma(0)$ for $\omega\ =\ 0$ is significantly different from each other.

%*******************************************************************************
%     Discussions & Conclusion
%*******************************************************************************
To interpret the above universal correction due to frequency $\omega$,
we recall the weak localization correction,
\begin{equation}
\delta \sigma = -\frac{e^2}{2 \pi^2 \hbar}
\ln{\frac{\tau_\phi}{\tau}},
\end{equation}
where $\tau_{\phi}$ is the dephasing time, and
$\tau$ the elastic scattering time.
The $\omega$ dependence of $\tau_{\phi}$ in the
kicked rotator problem\cite{fishman91} as well as that
in the one-dimensional  disordered system \cite{borgonovi95}
in the presence of noise
is estimated to be $\sim \omega^{-2/3}$.
Then the weak localization correction $\delta\sigma$~\cite{bergmann84,HLN}
is estimated to be
\begin{equation}
\delta \sigma =
 \frac{2}{3\pi}\frac{e^2}{h}\ln{\frac{\omega}{\omega_0}} .
\label{eqn:weaklocalization}
\end{equation}
The pre-factor $2/3\pi\ =\ 0.212\dots$ of the $\ln{\omega}$ term
agrees with the numerical calculation
shown in Fig. \ref{fig:universal},
which is universal and independent of the potential strength $W$.
Note that this $\omega$-dependence is also observed in
the case of low energy phonon scattering.\cite{note}

To justify our argument,
the condition $\tau\ll\tau_{\phi}$ is necessary.
This is why the conductivity deviates from the $\log \omega$ behavior
in large $\omega$ region,
where $\tau$ is the order of $\tau_{\phi}$.
We also see the deviation from the $\log \omega$ behavior
in a small $\omega$ region.
In this case, $\omega$ is too small for the potential to be changed
in a finite time where our simulation has been performed.

In conclusion, we have analyzed the weak localization effect
in the two-dimensional system with time-dependent random potentials
using the equation-of-motion method.
It has been shown numerically that the weak localization correction
to the conductivity due to the fluctuating potentials
takes universal value independent of the random potential strength $W$.
The correction term agrees with the formula
$\delta \sigma\ =\ \frac{2}{3\pi}\frac{e^2}{h}\ln{\frac{\omega}{\omega_0}}$,
proposed for the inelastic scattering by moving impurities.
Our results will open a new way to incorporate the dephasing mechanism
in numerical simulations.

%*******************************************************************************
%     Acknowlegments
%*******************************************************************************
One of the authors (T. N.) acknowledges the financial support from
the Proposal-Based Advanced Industrial Technology R\&D Program of
the NEDO.
Numerical calculations were performed on FACOM VPP500 in Supercomputer
Center, Institute for Solid State Physics, University of Tokyo.
%*******************************************************************************
%     References
%*******************************************************************************

%*******************************************************************************
%     Figure caption
%*******************************************************************************
\begin{figure}
\noindent
Fig. \ref{fig:1sample}:
An example of the second moment as the function of time for
$\omega\ =\ 10^{-3} V/\hbar,$ $W/V\ =\ 4\sqrt{2},$ and
$E_{\mbox{F}}/V\ =\ -1$. The solid line corresponds
to $D\ =\ 1.46a^2V/\hbar$.\\
Fig. \ref{fig:recursion}:
Conductivity obtained by the equation-of-motion
method (solid line) for $\omega\ =\ 0$. The dashed line corresponds to
the conductance calculated by the recursive Green's function method.\\
Fig. \ref{fig:universal}:
The calculated conductivity of the two-dimensional system with time
dependent impurity potential for (i) $W/V\ =\ 2\sqrt{2}$, (ii)$W/V\ =\ 4$,
and (iii) $W/V\ =\ 4\sqrt{2}$.
The solid lines indicate the corrections
$\delta\sigma\ =\ 0.232\frac{e^2}{h}\ln{\tilde{\omega}}$,
$\delta\sigma\ =\ 0.217\frac{e^2}{h}\ln{\tilde{\omega}}$,
and $\delta\sigma\ =\ 0.214\frac{e^2}{h}\ln{\tilde{\omega}}$,
respectively.
\end{figure}
\end{document}